\documentclass[twocolumn,showpacs,preprintnumbers,amsmath,amssymb]{revtex4}

\usepackage{graphicx}
\usepackage{dcolumn}
\usepackage{bm}

\begin{document}
\title{Deformable self-propelled particles}
\author{Takao Ohta }
 \email{takao@scphys.kyoto-u.ac.jp} 
\affiliation{Department of Physics, Graduate School of Science, Kyoto University, Kyoto 606-8502, Japan}
\author{Takahiro Ohkuma}
  \email{ohkuma@ton.scphys.kyoto-u.ac.jp} 
\affiliation{Department of Physics, Graduate School of Science, Kyoto University, Kyoto 606-8502, Japan}
\date{\today}

\begin{abstract}
A theory of self-propelled particles is developed in two dimensions assuming that the particles can be deformed from a circular shape when the propagating velocity is increased. A coupled set of equations in terms of the velocity and a tensor variable to represent the deformation is introduced to show that there is a bifurcation from a straight motion to a circular motion of a single particle. Dynamics of assembly of the particles is studied numerically where there is a global interaction such that the particles tend to cause an orientational  order.
\end{abstract}

\maketitle


Individual and collective dynamics of self-propelled objects are one of the fundamental problems in non-equilibrium physics. There have been a number of studies for many years based on both deterministic dynamics \cite{Levine, Chuang} and stochastic dynamics \cite{Vicsek, Schweitzer, Condat, Ebelingbook, Chate, Perunani}. Recently the hydrodynamics effects with and without internal degrees of freedom  have been investigated intensively \cite{Ramaswamy, Llopis, Aranson, Yeomans}. Formulation of a
droplet or a vesicle moving under the interaction with substrate has been studied  \cite{Durand, Pismen, Knobloch}.  Experiments of 
reaction-driven propulsion \cite{Nagayama, Yoshikawa} as well as  theory \cite{Baer} and computer simulations \cite{Golestanian, Kapral} have been carried out recently.
Excitable reaction diffusion systems have also exhibited self-propelled domains \cite{Mikhailov} and  have been analyzed theoretically \cite{Purwins, Ohta, Nishiura}.
 These studies are motivated, on one hand,  to understand  macroscopic self-organization far from equilibrium and, on the other hand, to clarify the mechanism of molecular machines in mesoscopic or nanoscopic length scales.

All of the above studies except for the model with internal degrees of freedom such as a dimmer model \cite{Yeomans} assume that the particle shape is unchanged during the motion. In reality, however, many self-propelled objects may change the shape depending on the velocity or the interaction with other objects. Biological systems such as living cells are one example \cite{Klages}. The excitable reaction diffusion system provides another example. The numerical results in ref. \cite{Mikhailov} clearly show that a two-dimensional excited pulse changes  its shape as the velocity increases. 

The purpose of the present Letter is to investigate the individual and the collective motions of self-propelled  deformable particles. It will be shown that a competition between a circular motion of individual particles and an orientational order of their shape due to a global interaction causes a rich variety of collective dynamics. 

\begin{figure}[hbpt]
\begin{center}
\includegraphics[width=.2\linewidth]{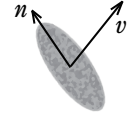}
\caption{Self-propelled particle with the velocity $\bm { v}$ and the unit normal $\bm {n}$ along the long axis.}
\label{vn}
\end{center}
\end{figure}

In order to represent the motion of a two-dimensional deformable particle with a coupling between the velocity of the center of the gravity and the deformation from a circular shape, we introduce two sets of variables. One is the velocity $ \bm {v} = (v_1, v_2)$ of the center of gravity and the other is a tensor $S_{\alpha \beta}$ with $(\alpha, \beta=1,2)$. Suppose that the deformation is weak and each particle deforms into an elliptical shape. We define a unit vector $\bm {n}$ along the long axis as shown in Fig. \ref{vn}. The tensor $S_{\alpha \beta}$ which is the same as the nematic order parameter \cite{Prost}  is given in terms of the components of $\bm { n}$ by
\begin{eqnarray}
S_{\alpha \beta} = s \left( n_{\alpha} n_{\beta} - \frac{1}{2} \delta_{\alpha \beta} \right)
\label{S}
\end{eqnarray}
which is normalized as Tr$S=0$ and the positive constant $s$ represents the degree of deformation from a circular shape. 

The time-evolution equations for  $ \bm {v}$ and $S_{\alpha \beta}$ can be derived by a symmetry argument as follows. First of all, the velocity should obey in its simplest case 
\begin{eqnarray}
\frac{d}{dt} v_{\alpha} &=& \gamma v_{\alpha} - | \bm{v}^{2} | v_{\alpha} - a S_{\alpha \beta} v_{\beta} 
\label{eqv}
\end{eqnarray}
where the repeated indices imply summation.
The constant $\gamma$ is assume to be positive. The first and the second terms are the same as those of an active Brownian particle~\cite{Schweitzer, Condat, Ebelingbook}. 
The combination $S_{\alpha \beta}v_\beta$ with a scalar constant $a$ as the last term of (\ref{eqv}) is the simplest way  to make a vector variable in terms of  $ \bm {v}$ and $S_{\alpha \beta}$.  Similarly,  by symmetry, a tensor $v_{\alpha} v_{\beta} - \frac{1}{2} | \bm{v}^{2} | \delta_{\alpha \beta}$ should enter into  the time-evolution equation for $S_{\alpha \beta}$
\begin{eqnarray}
 \frac{d}{dt} S_{\alpha \beta} &=& - \kappa S_{\alpha \beta} + b \left( v_{\alpha} v_{\beta} - \frac{1}{2} | \bm{v}^{2} | \delta_{\alpha \beta} \right)
\label{eqS}
\end{eqnarray}
where $\kappa > 0$ and $b$ are  constant. Note that (\ref{eqS}) satisfies Tr$S=0$.


By writing the velocity as $v_1= v \cos \phi$, $\ v_2= v \sin \phi $ and the unit normal as $n_1=\cos \theta$ and $n_2=\sin \theta$, eqs.  (\ref{eqv})  and (\ref{eqS}) are rewritten as
\begin{eqnarray}
\frac{d}{dt} v &=& v(\gamma - v^{2}) - \frac{1}{2} asv \cos 2(\theta - \phi) 
\label{ampv}\\
\frac{d}{dt} \phi&=&- \frac{1}{2} as \sin 2 (\theta - \phi) 
\label{phi}\\
\frac{d}{dt}s &=& -\kappa s+ v^{2}b \cos 2 (\theta -\phi) 
\label{ampS}\\
\frac{d}{dt} \theta &=& - \frac{v^{2} b}{2s} \sin 2 (\theta - \phi)
\label{theta}
\end{eqnarray}
where $v$ and $s$ should be positive.
From eqs. (\ref{phi}) and  (\ref{theta}), we obtain for $\psi=\theta - \phi$
\begin{eqnarray}
 \frac{d}{dt} \psi &=& -\frac{1}{2} \left (-as+\frac{bv^2}{s}\right)\sin2\psi
\label{psi}
\end{eqnarray}
It is noted that only the difference $\psi$ of the two angles enters into the time-evolution equations due to the spatial isotropy. 

The above set of equations has a stationary solution. Without loss of generality, we may assume that the velocity is directed along the $x$-axis, i.e., $\phi=0$. From eqs.  (\ref{theta})  and (\ref{phi}), one notes that the angle $\theta$ is given either by $\theta=0$ or by $\theta=\pi/2$ depending on the sign of $b$.
That is, when $b$ is positive, $\theta=0$, and the steady value of the amplitudes is given by $v_0^2=\gamma/(1+B)$ and $s_0=bv_0^2/\kappa$ where $B\equiv ab/(2\kappa)$. Therefore the long axis of the elliptical particle is parallel to the propagating direction. On the other hand, when $b$ is negative, the solution is given by  $\theta=\pi/2$, $v_0^2=\gamma/(1+B)$ and $s_0=-bv_0^2/\kappa$. In this case, the long axis is perpendicular to the velocity vector. In order to avoid a singular behavior of $v_0$ for $B \le -1$, we hereafter assume $ab > 0$ ($B > 0$).

\begin{figure}[hbpt]
\begin{center}
\includegraphics[width=1.0\linewidth]{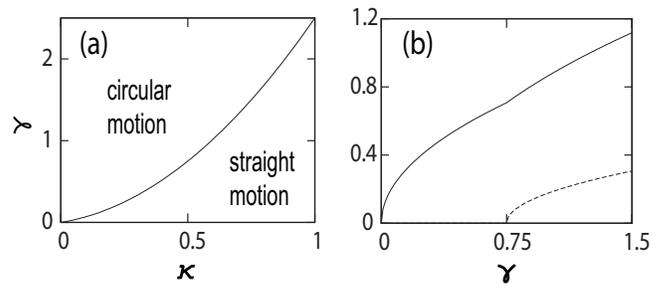}
\caption{(a) Stability diagram on the $\gamma-\kappa$ plane for $ab=0.5$. (b) The velocity $v$ (full line) and the frequency $\omega$ (broken line) as a function of $\gamma$ for $a=-1.0,b=-0.5$ and $\kappa=0.5$. The bifurcation occurs at $\gamma_c = 0.75$.}
\label{critical}
\end{center}
\end{figure}

The linear stability of the stationary solution can be examined by a standard method. It is readily shown that the linearized equations for the amplitudes  (\ref{ampv})  and (\ref{ampS}) are decoupled with those of the angle variables and any instability does not appear from the former. It turns out that the stability is solely determined by eq.  (\ref{psi}). The stationary solution is stable when the coefficient of rhs of eq.  (\ref{psi}) is negative. This gives us the condition that 
$abv_0^2 \le \kappa^2$
and hence  the threshold that
\begin{eqnarray}
\gamma=\gamma_c= \frac{\kappa^2 }{ab}+\frac{\kappa}{2} 
\label{gammac}
\end{eqnarray}
When $\gamma \ge \gamma_c$, the stationary straight motion becomes unstable.
 Note that this is valid for both $\theta=0$ and $\theta=\pi/2$. This bifurcation threshold is indicated on the $\gamma-\kappa$ plane in Fig. \ref{critical}. The instability occurs  when the velocity exceeds the critical value defined by
$v_c=\kappa /(ab)^{1/2}$.
We assume that the value of $v_c$ is sufficiently small. This is required because the deformation  is expected to be enhanced for larger velocity whereas the deviation from the circular shape in our model is expressed only by $S_{\alpha \beta}$ with the mode $2\theta$ ignoring the higher harmonics. 

\label{vomega}

Now we investigate the dynamics of a self-propelled particle when the steady solution $v_0$ and $s_0$ of a straight motion becomes linearly unstable.  Here we make an ansatz that there occurs a circular motion and explore the stability both numerically and analytically. To this end, we put $v=v_r$, $s=s_r$ and 
$\theta = \omega t+\zeta/2$
and 
$ \phi = \omega t$.
Substituting these into eqs.    (\ref{ampv}),  (\ref{phi}), (\ref{ampS}) and (\ref{theta}),  one obtains
after some algebra 
\begin{eqnarray}
v_r^2 =\gamma-\frac{\kappa}{2}
\label{vr}
\end{eqnarray}
and
$s_r^2 = bv_r^2/a$,
$\cos \zeta =\kappa/as_r$
and 
$\omega^2 =(ab/4)(v_r^2-v_c^2)$.
Equation  (\ref{vr}) requires that $\gamma \ge \kappa/2$ but it is found from eq. (\ref{gammac}) that this is automatically satisfied when $\gamma \ge \gamma_c$.
It should also be noted that this inequality is the condition for the existence of the phase $\zeta$.
The frequency $\omega$ continuously increases from zero at $v_r=v_c$. The radius of the circular orbit is given by $R=v_r/\omega$.
%
Comparing the stationary solution  $v_0^2=\gamma/(1 + B)$ with $v_r$ given by eq.  (\ref{vr}), one notes that $v$ is continuous at the bifurcation point $\gamma=\gamma_c$.
The $\gamma-$dependence of $v$ and $\omega$ is shown in Fig. \ref{critical}(b).

In order to study the dynamics near the bifurcation $\gamma=\gamma_c$ in more detail, we make a reduction of the variables. It is noted that the coefficient of eq. (\ref{psi}) vanishes at the bifurcation point. Therefore, the variable $\psi$ is slow near the stability threshold compared to other variables $v$ and $s$. This allows us to eliminate these variables by putting $dv/dt=ds/dt=0$ in eqs.  (\ref{ampv})  and (\ref{ampS}) so that eq. (\ref{psi}) becomes 
\begin{eqnarray}
 \frac{d}{dt} \psi &=& F(\psi)\nonumber \\
 & =& -\kappa\left[\frac{\gamma-\gamma_c+(\kappa/2-\gamma)(1-\cos^22\psi)}{-2\gamma_c+\kappa(1-\cos^22\psi)} \right]\tan 2\psi \nonumber \\
\label{psi2}
\end{eqnarray}
The function $F(\psi)$ is an odd function since both clock-wise and counter clock-wise circular motions are equally possible. It is readily verified  that  eq. (\ref{psi2}) for $-\pi/2 < \psi < \pi/2$ is monostable for $\gamma \le \gamma_c$ whereas it becomes bistable for $\gamma \ge \gamma_c$. Therefore this is a pitchfork bifurcation.


\begin{figure}[hbpt]
\begin{center}
\includegraphics[width=1.\linewidth]{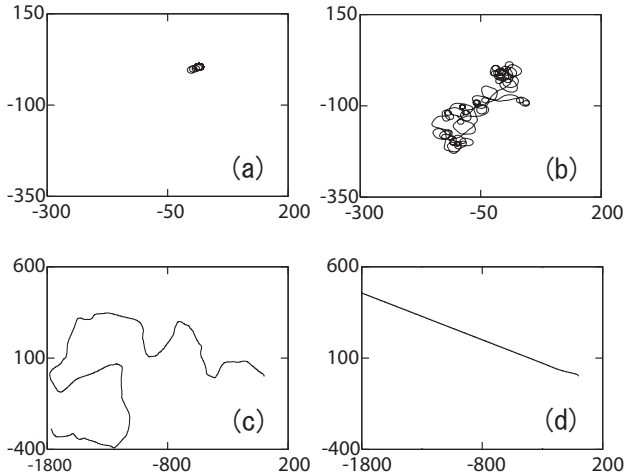}
\caption{Trajectory of a particle for (a) $K=\hat{K}/\kappa=0.14$ and for the time interval $\delta t=5000$, (b) $K=0.16$ and for $\delta t=5000$, (c) $K=0.28$ and for $\delta t=5000$, and (d) $K=0.32$ and for $\delta t>2000$. Note that (c) and (d) cover wider area compared with (a) and (b). 
}
\label{trajectory}
\end{center}
\end{figure}

The linear stability of the circular motion is studied numerically by evaluating the eigenvalues of the linearized equations of  (\ref{ampv}), (\ref{ampS})  and  (\ref{psi}). It turns out that the stability threshold 
coincides with the stability threshold of the straight motion given in Fig. \ref{critical}. This implies that in the region where the straight steady motion is stable, the circular motion is unstable and vice verse. 
 
 It is mentioned here that a model of self-propelled particle which exhibits a circular motion has been introduced quite recently by Teeffelen and Loewen  \cite{Loewen}. However their model does not have a bifurcation from a straight motion to a circular motion shown above.
 
Now we explore the collective motion of the deformable particles. We introduce a kind of a global coupling such that 
\begin{eqnarray}
 \frac{d}{dt} S^{(n)}_{\alpha \beta} &=& - \kappa S^{(n)}_{\alpha \beta} + b \left( v^{(n)}_{\alpha} v^{(n)}_{\beta} - \frac{1}{2}  (\bm{v}^{(n)})^{2}  \delta_{\alpha \beta} \right) \nonumber \\
 & &-  \hat{K}(S^{(n)}_{\alpha \beta}-\bar{S})
\label{eqS2}
\end{eqnarray}
where $\bar{S}=(1/N)\sum_nS^{(n)}_{\alpha \beta}$ and $\hat{K}$ is the coupling constant. The superscript $n$ indicates the $n$-th particle. The total number of the particles is denoted by $N$.  Equation  (\ref{eqS2}) should be coupled with eq. (\ref{eqv}) in which $v_{\alpha}$ and $S_{\alpha \beta}$ are replaced, respectively, by $S^{(n)}_{\alpha \beta}$ and $S^{(n)}_{\alpha \beta}$.
The last term in eq.  (\ref{eqS2}) with a positive value of $\hat{K}$ implies that the ellipsoidal particles tend to make an orientational oder. It is mentioned that an orientational order has been studied in self-locomoting rods which are not deformable \cite{Saintillan}.

\begin{figure}[hbpt]
\begin{center}
\includegraphics[width=1.0\linewidth]{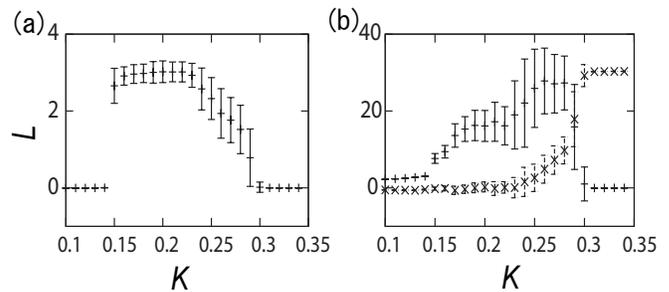}
\caption{(a) Lyapunov exponent as a function of the interaction strength $K$. (b) Diffusion constant $D$ (plus) and the coefficient $B$ (cross) in eq. (\ref{WD})  as a function of the interaction strength $K$. These are obtained from 100 independent runs. }
\label{diffusion}
\end{center}
\end{figure}

We have investigated the motion of the particles numerically in the situation such that individual particle undergoes a circular motion when $K=0$. The parameters are fixed as $a=-1.0, b=-0.5$, $\kappa=0.5$ and $\gamma=1.0$ whereas the interaction strength $K=\hat{K}/\kappa$ is varied. The number of the particles is chosen as $N=30$ in the most of numerical simulations.

Figure \ref{trajectory} displays  the trajectory of a particle for four different values of $K$. When $K$ is small as $K=0.14$, the motion is almost circular and localized as in Fig. \ref{trajectory}(a). However a transition to a delocalized state occurs around at $K=0.15$. A random drift motion appears as in  Fig. \ref{trajectory}(b). For larger values of $K$, a random but ballistic motion becomes dominant as shown in Fig. \ref{trajectory}(c). When $K$ exceeds 0.30, each particle moves straightly as displayed in  Fig. \ref{trajectory}(d). The direction of the velocity of each particle is completely random.

In order to analyze the above behavior of the collective motion, we have evaluated the Lyapunov exponent $L$ defined through the relation 
\begin{eqnarray}
\exp(L(t-t_0)/\tau) &=& \sum_{n=1}^{N} \left[ (\bm{v}^{(n)} (t) - \hat{\bm{v}}^{(n)} (t))^{2} \right. \nonumber \\
& & \left. - \det(S^{(n)} (t) - \hat{S}^{(n)} (t)) \right] 
\label{Lyapunov}
\end{eqnarray}
where $\tau=2 \pi / \omega =8\sqrt{2}\pi\approx 35.5$ for the present set of the parameter values.
Small perturbations of the order of $10^{-4}$ are introduced for the hat-marked variables at $t_0=3200$. The exponent $L$ is obtained by changing the value of $K$. The results are summarized in Fig. \ref{diffusion}(a). The Lyapunov exponent is almost zero for $K \le 0.14$ and $K \ge 0.3$ whereas it is definitely positive for $0.15 \le K \le 0.29$. Therefore we conclude that the motion is chaotic in this parameter regime. In Fig. \ref{diffusion}(a), the data of 100 independent runs are plotted. The vertical bars indicate the typical scatter of the data. The reason why the exponent is not negative but zero in the non-chaotic region is that there is a zero eigen-mode in the time-evolution equations due to the rotational invariance.

In order to quantify the distinct dynamics in these parameter regions, we have calculated the averaged mean square displacement for all 30 particles defined by
\begin{eqnarray}
W= \left[ \frac{1}{N} \sum_{n=1}^{N} \left( \int_{0}^{t} \bm{v}^{(n)} (t') dt' \right)^{2} \right]^{1/2} .
\label{W}
\end{eqnarray}
 Since there is a crossover from a random drift to a ballistic motion of a particle as is evident in Fig. \ref{trajectory}, we assume the time-dependence of $W$ for sufficiently large values of $t$ as
\begin{eqnarray}
W=2D(t/\tau)^{1/2} +Bt/\tau.
\label{WD}
\end{eqnarray}
The diffusion constant $D$ defined through the relation (\ref{WD}) is plotted in Fig. \ref{diffusion}(b) as a function of the interaction strength $K$. It is clearly seen that the diffusion constant starts to increase in the chaotic region and becomes zero for $K \ge 0.3$ whereas the ballistic motion becomes dominant and the value of $B$ increases for $K\ge 0.25$ as shown in Fig. \ref{diffusion}(b). The meaning of the diffusion constant is less appropriate for these large values of $K$.

In summary, we have introduced and studied a kinetic model for deformable self-propelled particles. It is found that an isolated particle exhibits a bifurcation such that a straight motion becomes unstable and a circular motion appears. It is noted that this is not an ordinary Hopf bifurcation 
 and that the frequency of the circular motion increases continuously from zero at the bifurcation point. As indicated by eq. (\ref{psi2}), the bifurcation is a pitchfork type accompanied with a zero mode due to the spatial isotropy.

Assembly of the deformable particles exhibits complex dynamics where a global interaction is introduced in such a way that the particles deformed in an elliptical shape tend to have an orientational order. In the situation that each particle undergoes the circular motion if the global interaction is absent, three phases appear by increasing the interaction strength, the localized phase, delocalized chaotic phase and the phase of the ballistic motion. The final ballistic phase can be understood easily. When the direction of the velocity and hence the orientation of each particle is random, the average $\bar{S}$ is zero in eq.  (\ref{eqS2}) so that this equation is decoupled for each particle. The damping constant $\kappa$ is renormalized as $\kappa+\hat{K}$ and hence, for large values of $\hat{K}$, this is in the phase of straight motion in Fig.  \ref{critical}(a). It is expected qualitatively that the chaotic dynamics in the intermediate parameter region originates from a competition between the circular motion of individual particles and the tendency of the orientation order for increasing the interaction strength.  
Finally, it is mentioned that our preliminary numerical simulations have verified that the above property of the complex collective dynamics is preserved when an attractive interaction whose strength is proportional to  the distance of a pair of particles.  The full account of these results together with the theoretical analysis will be published elsewhere in the neat future.

This work was supported 
 the Grant-in-Aid for the priority area "Soft Matter Physics" 
from the Ministry of Education, Culture, Sports, Science and Technology (MEXT) of Japan.


\begin{thebibliography}{1}

\bibitem{Levine}
H. Levine, W-J. Rappel and I. Cohen, Phys. Rev. E {\bf 63}, 017101 (2000).
\bibitem{Chuang}
Y-l, Chuang, M. R D'Orsogna, D. Marthaler, A. L. Bertozzi and L. S. Chayes, Physica D{\bf 232}, 33 (2007).

\bibitem{Vicsek}
T. Vicsek et. al, Phys. Rev. Lett.  {\bf 75}, 1226 (1995).
\bibitem{Schweitzer}
F. Schweitzer, W. Ebeling and B. Tilch, Phys. Rev. Lett.  {\bf 80}, 5044 (1998).
\bibitem{Condat}
C. A. Condat and G. J. Sibona, Physica D  {\bf 168-169}, 235 (2002).
\bibitem{Ebelingbook}
W. Ebeling and I. M. Sokolov, {\it Statistical Thermodynamics and Stochastic Theory of Nonequilibrium Systems} (World Scientific Publishing, London, 2005) 
\bibitem{Chate}
G. Gregoire and H. Chate, Phys. Rev. Lett.  {\bf 92}, 025702 (2004).
\bibitem{Perunani}
F. Peruani and L. G. Morelli, Phys. Rev. Lett.  {\bf 99}, 010602 (2007).

\bibitem{Ramaswamy}
R. A. Simha and S. Ramaswamy, Phys. Rev. Lett.  {\bf 89}, 058101 (2002).
\bibitem{Llopis}
I. Llopis and I. Pagonabarraga, Europhys. Lett.  {\bf 75},  999 (2006).
\bibitem{Aranson}
I. S. Aranson et. al, Phys. Rev. E {\bf 75}, 040901(R) (2007).
\bibitem{Yeomans}
G. P. Alexander and J. M. Yeomans, Europhys. Lett.  {\bf 83},  34006 (2008).


\bibitem{Durand}
I. Durand, P. Joenson, C. Mishbah, A. Valance and K. Kassner, Phys. Rev. E{\bf 56}, R3776 (1997).

\bibitem{Pismen}
L. M. Pismen, Phys. Rev. E {\bf 74}, 041605 (2006).
\bibitem{Knobloch}
U. Thiele and E. Knobloch, Phys. Rev. Lett.  {\bf 97}, 204501 (2006).

\bibitem{Nagayama}
M. Nagayama et al,  Physica D  {\bf 194}, 151 (2004).
\bibitem{Yoshikawa}
Y. Sumino and K. Yoshikawa,  Chaos  {\bf 18}, 026106 (2008).
\bibitem{Baer}
K. John et al, Eur. Phys. J. E  {\bf 18}, 183 (2005).
\bibitem{Golestanian}
R. Golestanian et al, Phys. Rev. Lett.  {\bf 94}, 220801 (2005).
\bibitem{Kapral}
Y-G. Tao and R. Kapral, J. Chem. Phys. {\bf 128}, 164518 (2008).



\bibitem{Mikhailov}
K. Krisher and A. Mikhailov, Phys. Rev. Lett.  {\bf 73}, 3165 (1994).
\bibitem{Purwins}
M. Or-Guil et al, Phys. Rev. E {\bf 57}, 6432 (1998).
\bibitem{Ohta}
T. Ohta, Physica D  {\bf 151}, 61 (2001).
\bibitem{Nishiura}
Y. Nishiura et al, Chaos  {\bf 15}, 047509 (2005).


\bibitem{Killich}
T. Killich et al, J. Cell Sci.  {\bf 106}, 1005 (1993).
\bibitem{Klages}
P. Dieterich et al, Proc. Natl. Acad. Sci. USA  {\bf 105}, 459 (2008).




\bibitem{Prost}
P. G. de Gennes and J. Prost,  {\it The Physics of Liquid Crystals} (Oxford University Press, Oxford, 1993) 


\bibitem{Saintillan}
D. Saintillan and M. J. Shelley, Phys. Rev. Lett.  {\bf 99}, 058102 (2007).


\bibitem{Loewen}
S. van Teeffelen and H. Loewen, Phys. Rev. E {\bf 78}, 020101(R) (2008).

\end{thebibliography}
\end{document}